\documentclass[showpacs,preprintnumbers,amsmath,amssymb,10pt]{revtex4}

\usepackage{dcolumn}% Align table columns on decimal point
\usepackage{bm}% bold math

\usepackage{indentfirst}
\usepackage{graphicx}

\def\Vec#1{{\bf #1}}

\newcommand{\p}{\perp}
\newcommand{\ssh}{\!\!\!/}

\newcommand{\bi}{\begin{itemize}}
\newcommand{\ei}{\end{itemize}}
\newcommand{\be}{\begin{equation}}
\newcommand{\ee}{\end{equation}}
\newcommand{\ba}{\begin{eqnarray}}
\newcommand{\ea}{\end{eqnarray}}

\begin{document}

%\preprint{USM-TH-229}

\title{Azimuthal angle dependence of di-jet production in unpolarized hadron scattering}

\author{Zhun Lu}
\author{Ivan Schmidt}\affiliation{Departamento de F\'\i sica,
Universidad T\'ecnica Federico Santa Mar\'\i a, Casilla 110-V,
Valpara\'\i so, Chile\\ and Center of Subatomic Physics, Valpara\'\i
so, Chile}

\begin{abstract}
We study the azimuthal angular dependence of back-to-back di-jet
production in unpolarized hadron scattering $H_A+H_B \to J_1 + J_2
+X$, arising from the product of two Boer-Mulders functions, which
describe the transverse spin distribution of quarks inside an
unpolarized hadron. We find that when the di-jet is of two identical
quarks ($J_q+J_q$) or a quark-antiquark pair ($J_q+J_{\, \bar{q}}$),
there is a $\cos \delta \phi$ angular dependence of the di-jet, with
$\delta \phi=\phi_1-\phi_2$, and $\phi_1$ and $\phi_2$ are the
azimuthal angles of the two individual jets. In the case of
$J_q+J_q$ production, we find that there is a color factor
enhancement in the gluonic cross-section, compared with the result
from the standard generalized parton model. We estimate the $\cos
\delta \phi$ asymmetry of di-jet production at RHIC, showing that
the color factor enhancement in the angular dependent of $J_q+J_q$
production will reverse the sign of the asymmetry.
\end{abstract}

\pacs{12.38.Bx, 13.85.Ni, 13.88.+e}

\maketitle

\section{Introduction}
 The role of transverse momentum dependent (TMD)
distributions~\cite{mulders,bm} in high energy physics has received
a great deal of attection recently. This major interest in TMD
distributions relies on the belief that they are responsible for
various azimuthal asymmetries observed in different single particle
inclusive processes involving at least two hadrons. Examples are the
single transverse spin asymmetry (SSA) measured in semi-inclusive
deeply inelastic scattering (SIDIS)~\cite{Airapetian:2004tw,compass}
and inclusive pion production in hadron
collision~\cite{e704_91,star2004}, as well as large $\cos 2 \phi$
anomalous asymmetry in the Drell-Yan
process~\cite{na10,conway,e866}. It is found that leading-twist
time-reversal-odd ($T$-odd) distributions play an essential role in
these asymmetries, i.e. the Sivers function~\cite{sivers} can
account for the SSA, while the Boer-Mulders function~\cite{bm,boer}
can explain the large $\cos 2 \phi$ anomaly. A key ingredient of
$T$-odd distributions is the path-ordered exponential contained in
their definition, the so called Wilson line or
gauge-link~\cite{collins02,bjy02,bmp03} due to initial-state or
final-state interactions~\cite{bhs02} between the active partons and
the spectator system. For example, final state interactions are
related to a future pointing Wilson line (denoted as
$\mathcal{U}^{[+]}$) for $T$-odd distributions, connected to the SSA
in SIDIS processes. On the other hand, the study of Drell-Yan
processes shows that initial-state interactions give rise to a
past-pointing Wilson line ($\mathcal{U}^{[-]}$), which predicts a
minus sign difference of the Sivers
functions~\cite{collins02,bmp03,bhs02} and correspondingly the SSA
in the Drell-Yan process compared to the SIDIS process. Based on
these studies, TMD Factorization in SIDIS, Drell-Yan and $e^+ e^-$
annihilation has been established~\cite{Ji:2004xq,collins2004},
within the allowance of a sign reversal for $T$-odd functions in
different processes.

Recently further
studies~\cite{bv04,bomhof04,bacchtta05,bomhof06,bomhof07,cq07,
qvy07, vy07} have been performed on back to back di-jet or di-hadron
production in hadron-hadron collisions. Unlike in SIDIS and
Drell-Yan processes, in which there is only either final-state
interactions or initial state interactions, in hadron-hadron
collisions both initial-state and final-state interactions
contribute to the process. The result is that there is a more
complicated form of the gauge-link for TMD distribution functions in
hadron-hadron collisions than in SIDIS or Drell-Yan processes. For
example the loop structure
$\mathcal{U}^{[\Box]}=\mathcal{U}^{[-]^\dag}\mathcal{U}^{[+]}$ will
appear in the corresponding gauge-link~\cite{bomhof04}. Further
studies~\cite{bomhof06} show that in hadron-hadron collisions the
form of the gauge-link depends even on the hard partonic process.
The impact of these nontrivial gauge-links on phenomenology,
especially on SSA, has been studied in back-to-back hadron pair or
di-jet production~\cite{bacchtta05,bmvf07,bomhof07}, and photon-jet
productions~\cite{bacchetta07}. In Refs.~\cite{collins07,vy07}, the
authors show that in second order of gluon exchange from initial-
and final-state interaction, the standard TMD factorization is
violated in back-to-back hadron (jet) pair production, i.e., the TMD
distributions need to be modified in a process-dependent way (See
also Ref.~\cite{bmnpb08}, where the universal-breaking part of the
TMD distributions has been given).

As shown in Refs.~\cite{bacchtta05,bomhof07}, a consequence of
nontrivial gauge-links is that in hadron-hadron scattering the
parton distribution (and fragmentation) functions are convoluted
with gluonic pole cross-sections rather than the standard basic
parton cross-sections. Compared to the basic parton cross-sections,
there is a gluonic pole factor $C_G$ along with the gluonic
cross-section, which arises from the $T$-odd effect contributed by
the multiple initial- and final-state interactions mediated by gluon
exchanges. The complete gluonic pole cross-section encountered in
SSA in hadron-hadron scattering have been given in
Ref.~\cite{bacchtta05,bomhof07} for hadron pair, di-jet and
hadron-jet production, and in Ref.~\cite{bacchetta07} for photon-jet
production. In this paper, we will study the azimuthal angle
dependence of back-to-back di-jet produced in unpolarized
hadron-hadron scattering. Like the SSA in di-jet production, where
one incident hadron is transversely polarized, there is also
interesting phenomenology in the case of di-jet production in
unpolarized hadron-hadron scattering. First,  one realizes that
T-odd TMD distributions, especially the Boer-Mulders function,
produces also azimuthal asymmetries in unpolarized hadron
scattering, similar to the case of the $\cos 2 \phi$ asymmetry in
unpolarized Drell-Yan processes. A recent example is the analysis of
that asymmetry of photon-jet production in unpolarized hadron
scattering, given in Ref.~\cite{bmp07}. Therefore the study will
provide more knowledge on the TMD distributions which play a role in
the process. Second, it is interesting to investigate the effect of
multiple initial- and final-state interactions on the cross-section
of di-jet production in unpolarized hadron scattering, as has been
done in single transversely polarized hadron scattering. In
unpolarized hadron scattering the azimuthal asymmetry can be
produced by two $T$-odd distributions, which will make the gluonic
pole factor more complicated. Therefore a quantitative study of the
effect of multiple initial- and final-state interaction on
unpolarized hadron scattering will shed light on the QCD dynamics
that is present in azimuthal asymmetries.

\section{Angular dependence of di-jet production in unpolarized hadron scattering}

The process we study is
 \be H_A(P_A) + H_B(P_B)\rightarrow
J_1(P_1)+J_2(P_2) +X, \ee where the two incident hadrons $H_A$ and
$H_B$ are unpolarized. The momenta of these initial hadrons are
denoted by $P_A$ and $P_A$, and those of the di-jet by $P_1$ and
$P_2$. We are interested in the kinematical region where the
transverse momenta of the di-jet, $P_{1 \p}$ and $P_{2 \p}$, have
similar size and are almost back to back
\begin{equation}
\boldsymbol{P}_{1 \p} \approx -\boldsymbol{P}_{2 \p},
|\boldsymbol{P}_{1 \p}| \approx |\boldsymbol{P}_{2 \p}| \approx
P_{\p},
\end{equation}
and where the total transverse momentum of the di-jet
$\boldsymbol{q}_\p=\boldsymbol{P}_{1 \p} + \boldsymbol{P}_{2 \p}$ is
much smaller than $\boldsymbol{P}_\p$: $q_\p = |\,\boldsymbol{q}_\p|
\ll P_\p$. Therefore the process is sensitive to the intrinsic
transverse momenta of the partons inside the hadron. We label the
azimuthal angles of $\boldsymbol{P}_{1\p}$ and $P_{2\p}$ as $\phi_1$
and $\phi_2$, respectively. We also define $\delta \phi=\pi
-(\phi_2-\phi_1)$, which describes the imbalance of the di-jet from
a back to back configuration.

\begin{figure}

\begin{center}
\scalebox{0.63}{\includegraphics*{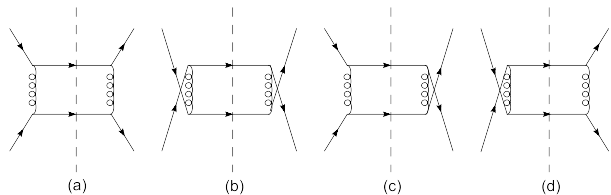}}\caption{\small
partonic diagrams contributing to di-jet production $H_A + H_B \to
J_q+J_q+X$. } \label{basicfig}
\end{center}

\end{figure}

The cross-section of di-jet production can be calculated from the
general form
 \begin{equation}
d\sigma = \frac{1}{2s}|\mathcal{M}|^2 \frac{d^3 P_1}{(2\pi)^3
2E_{P_1}} \frac{d^3 P_2}{(2\pi)^3 2E_{P_2}},
\end{equation}
where the amplitude square is expressed as a convolution of the
parton-parton to di-jet hard amplitudes and the correlation
functions~\cite{bacchtta05}:
\begin{eqnarray}
|\mathcal{M}|^2 &=& \int dx_1 d^2 \boldsymbol{k}_{1\p} dx_2 d^2
\boldsymbol{k}_{2\p}
(2\pi)^4  \nonumber\\
& \times & \delta ^4 (k_1+k_2-P_1-P_2)
\textrm{Tr}\{\Phi(x_1,\boldsymbol{k}_{1\p}^2)\Phi(x_2,\boldsymbol{k}_{2\p}^2)\nonumber\\
&\times & H(k_1,k_2,P_1,P_2)
H^\star(k_1,k_2,P_1,P_2)\}.\label{convolute}
\end{eqnarray}
In the above equation we denote the momenta of the initial partons
as $k_1$ and $k_2$, respectively.

The Mandelstam variables of the partonic process are defined as \ba
\hat{s}&=&(k_1+k_2)^2,\\
\hat{t}&=&(k_1-P_1)^2,\\
\hat{u}&=&(k_1-P_2)^2, \ea
which satisfy the relation \be
\frac{\hat{t}}{\hat{s}}=-y = -\frac{1}{e^{(\eta_1-\eta_2)}+1}, ~~~~~
\frac{\hat{u}}{\hat{s}} =-(1-y),
\end{equation}
where $\eta_{1/2}$ are the pseudorapidities of the jets.

\begin{figure}

\begin{center}
\scalebox{0.575}{\includegraphics*[171,566][256,698]{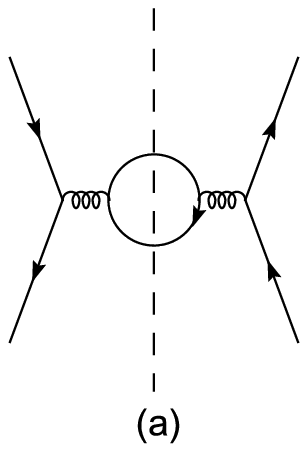}}
\scalebox{0.65}{\includegraphics*[160,582][237,698]{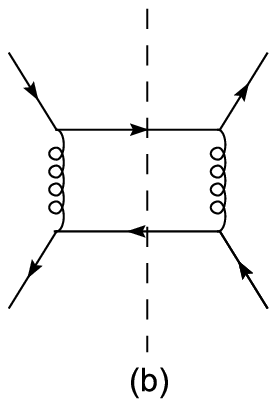}}
\scalebox{0.65}{\includegraphics*[160,582][237,698]{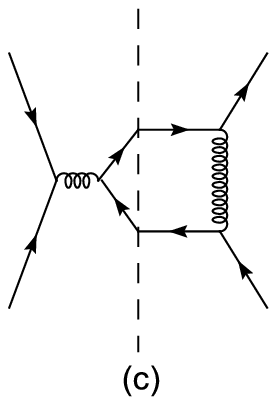}}
\scalebox{0.575}{\includegraphics*[171,565][256,698]{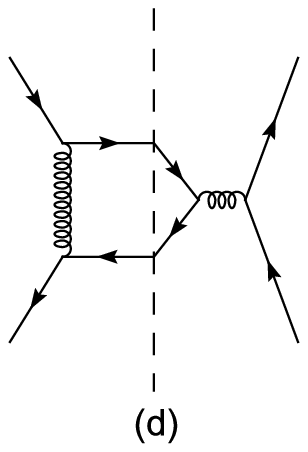}}
\caption{\small Partonic diagrams contributing to di-jet production
$H_A + H_B \to J_q+J_{\,\bar{q}}+X$ through $q \bar{q} \to q
\bar{q}$.} \label{qqbarfig}
\end{center}

\end{figure}

Since we will only consider unpolarized scattering, the correlation
matrix element (Soft part) for the unpolarized hadron can be simply
decomposed as~\cite{mulders,bm}
 \be
\Phi(P,x,\boldsymbol{k}_\p^2)=\frac{1}{2}\,\left
[f_1(x,\boldsymbol{k}_\p^2)P\!\ssh+\frac{
h_1^\p(x,\boldsymbol{k}_\p^2)}{M}k_\p^\mu P^\nu
\sigma_{\mu\nu}\,\right], \label{correlation}\ee
 where $f_1(x,\boldsymbol{k}_\p^2)$ is the
unpolarized TMD quark distribution, and
$h_1^\p(x,\boldsymbol{k}_\p^2)$ is the Boer-Mulders function
describing the transverse spin distribution of a quark in an
unpolarized hadron~\cite{bm}.

\begin{figure}[b]

\begin{center}
\scalebox{0.65}{\includegraphics*{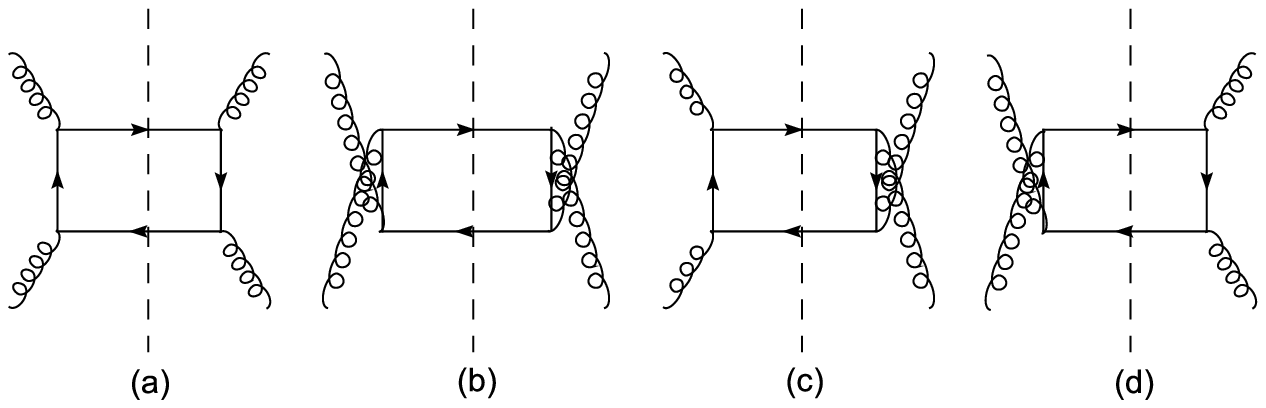}}\caption{\small
Partonic diagrams contributing to di-jet production $H_A + H_B \to
J_q+J_{\,\bar{q}}+X$ through $gg \to q \bar{q}$.} \label{qqbarfig2}
\end{center}

\end{figure}

In di-jet production, the transverse momenta of the particles in the
di-jet are measured in order to probe the corresponding parton
transverse momentum, and in this sense the azimuthal angle
dependence of the di-jet is of most interest. The cross-section of
unpolarized di-jet production contains both an azimuthal angle
independent part and an azimuthal angle dependent part. For the
azimuthal angle independent part, all possible partonic scattering
subprocesses can contribute to the process, while only the
unpolarized TMD distributions $f_1(x,p_T^2)$ does. On the other
hand, for the azimuthal angle dependent part, the Boer-Mulders
function $h_1^\p$ is involved, and since $h_1^\p$ is a chiral-odd
TMD function, only partonic processes that flip the helicity of the
quark will contribute. The possible partonic processes are $q (k_1)
+ q (k_2) \to q (P_1) + q (P_2)$ and $q (k_1) + \bar{q} (k_2) \to q
(P_1) + \bar{q} (P_2)$ (or $q^\prime(P_1)+\bar{q}^\prime (P_2))$.
Since the contribution of $h_1^{\p g}$ is power suppressed, we will
not consider the gluon contribution to the azimuthal angle asymmetry
in unpolarized di-jet production. Therefore we will specifically
study the azimuthal angular dependence of di-jet in unpolarized $H_A
+ H_B \to J_q+J_q+X$ and in $H_A + H_B \to J_q+J_{\bar{q}}+X$.
However, since quark-channels make only a small contribution to jet
rates (unless $P_\perp$ is very large) this suppression might be
compensated by the abundance of gluons and large scattering cross
sections for gluons.

\subsection{\bf Process $H_A + H_B \to J_q+J_q+X$}

For the process $H_A + H_B \to J_q+J_q+X$, the corresponding
partonic subprocess is $q \,q \to q \,q$, and the lowest-order cut
diagrams are shown in Fig.~\ref{basicfig}. According to
Eq.~(\ref{convolute}), one can therefore arrive at the angular
independent cross-section given by:
\begin{widetext}
\begin{eqnarray}
\frac{d\sigma^O[J_q \, J_q]}{d^2 \boldsymbol{P}_{1\p}d^2
\boldsymbol{P}_{2\p} d\eta_1 d\eta_2}&=& \frac{\alpha_s^2}{s\hat{s}}
\left \{\frac{N^2-1}{2N^2}\left
(\frac{1+(1-y)^2}{y^2}+\frac{1+y^2}{(1-y)^2}\right
)-\frac{(N^2-1)}{N^3}\frac{1}{y(1-y)} \right \}\nonumber \\
& \times & \sum_q e_q^2\int d^2 \boldsymbol{k}_{1\p} d^2
\boldsymbol{k}_{2\p} \delta^2
(\boldsymbol{k}_{1\p}+\boldsymbol{k}_{2\p}-\boldsymbol{P}_{1\p}-\boldsymbol{P}_{2\p})
f_1^q \,(x_1,\boldsymbol{k}_{1\p}^2)\,
f_1^q\,(x_2,\boldsymbol{k}_{2\p}^2),
 \end{eqnarray}
 \end{widetext}
with $\hat{s}=P_\p^2/(y(1-y))$. The momentum fractions $x_1$ and
$x_2$ can be expressed as
\begin{equation}
x_1=\frac{P_\p}{\sqrt{s}}(e^{\eta_1}+e^{\eta_2}), ~~~~
x_2=\frac{P_\p}{\sqrt{s}}(e^{-\eta_1}+e^{-\eta_2}).
\end{equation}
 After integrating over
$P_{2\p}$ and $\phi_1$, this leads to the expression
\begin{eqnarray}
& &\frac{d\sigma^O[J_q \, J_q]}{d P_{\p}^2 d\eta_1 d\eta_2}=
\frac{\pi}{s\hat{s}} \left \{\frac{N^2-1}{2N^2}\left
(\frac{1+(1-y)^2}{y^2}+\frac{1+y^2}{(1-y)^2}\right )\right.\nonumber
\\
& & \left. -\frac{(N^2-1)}{N^3}\frac{1}{y(1-y)} \right \} \sum_q
e_q^2
 f_1^q \,(x_1)\,
f_1^q\,(x_2), \label{csqqo}
\end{eqnarray}
 and the parton transverse momentum
integration is de-convoluted by the $\delta$-function. In the above
equation we have used $d^2 \boldsymbol{P}_{1\p} = \frac{1}{2} d
\boldsymbol{P}_{1\p}^2 d \phi_1 \approx \frac{1}{2} d P_{\p}^2 d
\phi_1$.

Our purpose is to study the azimuthal angle dependence of unpolarized
di-jet production. We recall that similar studies of angular
dependence have been performed for the case of the unpolarized
Drell-Yan process~\cite{boer}. Due to the correlation between the
quark transverse spin and transverse momentum, the product of two
Boer-Mulders functions $h_1^\p \times h_1^\p$ can give rise to large
$\cos 2 \phi$ angular asymmetries of dilepton production in
unpolarized Drell-Yan processes~\cite{boer,bbh03,lm05}. One can
expect that the same mechanism can produce non-zero angular
dependence of di-jets in unpolarized hadronic scattering. Actually a
recent study~\cite{bmp07} on photon-jet hadronic production shows
that $h_1^\p$ gives an angular dependence very similar to the $\cos
2 \phi$ asymmetry in Drell-Yan processes. We will follow this line
to study the azimuthal angle dependence of unpolarized di-jet
production.

Since $h_1^\p$ is a chiral-odd function which flips the helicity between
the incoming and ongoing quarks, among the diagrams
shown in Fig.~\ref{basicfig}, only Fig.~\ref{basicfig}c and
Fig.~\ref{basicfig}d can couple with $h_1^\p$ and consequently
contribute to the angular dependence of $J_q + J_q$ production. Thus
according to Eqs.~(\ref{convolute}) and (\ref{correlation}), the
contribution of $h_1^\p$ to the cross-section of unpolarized $J_q +
J_q$ can be expressed as
\begin{widetext}
\begin{eqnarray} & &\frac{d\sigma^A[J_q \, J_q]}{d^2 \boldsymbol{P}_{1\p}d^2
\boldsymbol{P}_{2\p} d\eta_1 d\eta_2} = \frac{4 C_{qq}
\alpha_s^2}{M^2s \boldsymbol{P}_\p^4} \left \{ y(1-y)\right \}\sum_q
e_q^2\int d^2 \boldsymbol{k}_{1\p} d^2 \boldsymbol{k}_{2\p} \delta^2
(\boldsymbol{k}_{1\p}+\boldsymbol{k}_{2\p}-\boldsymbol{P}_{1\p}-\boldsymbol{P}_{2\p}) \nonumber \\
& \times & \bigg{(}(\boldsymbol{k}_{1 \p} \cdot
\boldsymbol{k}_{2\p}) (\boldsymbol{P}_{1 \p} \cdot \boldsymbol{P}_{2
\p})-( \boldsymbol{k}_{1\p} \cdot \boldsymbol{P}_{1\p})
(\boldsymbol{k}_{2 \p} \cdot \boldsymbol{P}_{2 \p})-(
\boldsymbol{k}_{1\p} \cdot \boldsymbol{P}_{2\p})( \boldsymbol{k}_{2
\p} \cdot \boldsymbol{P}_{1 \p})\bigg{)}
 h_1^{\p q} \,(x_1,\boldsymbol{k}_{1\p}^2)\, h_1^{\p
q}\,(x_2,\boldsymbol{k}_{2\p}^2), \label{csqq}
 \end{eqnarray}
 \end{widetext}
where $C_{qq}$ is the color factor which will be determined below.
This result is very similar to the one obtained in photon-jet
production, and given in Ref.~\cite{bmp07}, where a $\cos 2 \phi$
angular dependence of photon-jet production is identified. Here we
will use a weighting procedure on Eq.~(\ref{csqq}). The advantage is
that the integrations over $\boldsymbol{k}_{1\p}$ and
$\boldsymbol{k}_{2\p}$ can be deconvoluted. We find that using the
weighting function $\mathcal{W}=\frac{ P_\p^2 \cos \delta
\phi}{M^2}$ and integrating the cross-section in Eq.~(\ref{csqq})
over $\boldsymbol{P}_{2\p}$, with $\delta \phi=\pi-
(\phi_2-\phi_1)$, we can arrive at a simpler form:
\begin{eqnarray}
&&\frac{d\sigma^{\mathcal{W}}[J_q \, J_q]}{d \boldsymbol{P}_{\p}^2
d\eta_1 d\eta_2} = \int d \phi_1 d^2 P_{2\p}  \frac{ P_\p^2 \cos
\delta \phi}{M^2} \nonumber \\
& \times & \frac{d\sigma^A[J_q \, J_q]}{d^2 \boldsymbol{P}_{1\p}d^2
\boldsymbol{P}_{2\p} d\eta_1 d\eta_2} \nonumber \\
& = & \frac{16 \pi C_{qq} \alpha_s^2 }{s \boldsymbol{P}_\p^2}
\,y(1-y) \, \sum_q e_q^2h_1^{\p,(1)}(x_1) \,
h_1^{\p,(1)}(x_2),\nonumber \\
\label{csqq3}
\end{eqnarray}
where $h_1^{\p (1)}(x)$ is the first $\boldsymbol{k}_\p^2$-moment of
$h_1^\p(x,\Vec k_T^2)$:
\begin{equation}
h_1^{\p (1)}(x) = \int d^2\boldsymbol{k}_\p
\frac{\boldsymbol{k}_\p^2}{2M^2} h_1^\p(x,\boldsymbol{k}_\p^2)
\end{equation}
Eq.~(\ref{csqq3}) shows that the angular dependent part of the
di-jet production cross-section can be obtained by a $\cos \delta
\phi$-moment of the total cross-section.

Now we will calculate the color factor $C_{qq}$ appearing in
Eq.~(\ref{csqq3}). There are two sources which contribute to
$C_{qq}$. One is the color factor from the hard scattering
subprocess. For the diagram in Fig.~\ref{basicfig}c, this color
factor is $-\frac{N^2-1}{4N^3}$, which is the standard result of the
generalized parton model. The other source is the special factor due
to the presence of $T$-odd distributions, the so call gluonic pole
factor~\cite{bacchtta05,bomhof07}. The key ingredients corresponding
to this factor are the initial- and final-state interactions (the
gluon exchanges before and after the hard scattering), which lead to
the gauge-links appearing in the definition of TMD T-odd
distributions. The simplest example is that every $T$-odd
distribution function will contribute a factor of $+1$ in SIDIS and
$-1$ in Drell-Yan process, due to the final- and initial-state
interactions, which correspond to  the future- and past-pointing
gauge links, respectively. In the case of hadron scattering the
situation is more involved, since the initial- and final- state
interactions are mixed, which makes the determination of gluonic
factors not straightforward. Systematic study of the gauge-links
appearing in hadron scattering has been performed in
Refs.~\cite{bacchtta05,bomhof07}. Here we will follow the approach
presented in Refs.~\cite{bacchtta05,bomhof07} to calculate the
gluonic pole factor coming from the product of two $h_1^\p$. The
general quark correlator with gauge link $\mathcal{U}$ is given by:
\begin{eqnarray}
\Phi^{[\mathcal{U}]}(x,\boldsymbol{k}_\p)& = & \int d\xi^- \int
\frac{d^2\boldsymbol{\xi}_\p}{(2\pi)^2} e^{ip \cdot \xi} \nonumber
\\
&\times & \langle
PS|\bar{\psi}(0)\,\mathcal{U}(0,\xi)\,\psi(\xi)|PS\rangle
\end{eqnarray}
This gauge-link is process dependent and also depends on the hard
subprocess. As we mentioned before, the subprocess shown in
Fig.~\ref{basicfig}c can couple with $h_1^\p$ and contributes to the
angular dependence of di-jet in hadron scattering. The gauge link
for the quarks which participate in this hard partonic process (the
interference of $t$ and $u$ channel) can be written
as~\cite{bacchtta05}:
\begin{eqnarray}
\mathcal{U}_{qq}^{[tu^*]} = \mathcal{U}_{qq}^{[tu^*]} &=&
\frac{2N^2}{N^2-1}\frac{\mathrm{Tr}(\mathcal{U^{[\Box]})}}{N}\,\mathcal{U}^{[+]}
\nonumber\\
&& -\frac{N^2+1}{N^2-1}\,\mathcal{U}^{[\Box]}\,\mathcal{U}^{[+]},
\label{wl1}
\end{eqnarray}
where
\begin{equation}
\mathcal{U^{[\Box]}}=\mathcal{U}^{[-] \dag}\mathcal{U}^{[+]}
\end{equation}
is the gauge link loop, and $\mathcal{U}^{[+]}$ and
$\mathcal{U}^{[-]}$ are the path-ordered future and past pointing
Wilson lines. The two terms appearing in Eq.~(\ref{wl1}) correspond
to the contributions from different color flows of the $q q \to q q$
process in the $tu^*$ channel. The convolution of the soft parts in the
decomposition of two color flows is therefore
\begin{eqnarray}
\Phi^{[\mathcal{U}]}\otimes \Phi^{[\mathcal{U}^\dag]} & \sim &
\frac{2N^2}{N^2-1}\Phi^{[(\Box)+]}\otimes
\Phi^{[(\Box)+^\dag]}\nonumber \\
&&-\frac{N^2+1}{N^2-1}\Phi^{[\Box +]}\otimes \Phi^{[\Box +^\dag]},
\end{eqnarray}
where $(\Box)+$ and $\Box +$ are the shortcuts of the basic gauge
links
$\frac{\mathrm{Tr}(\mathcal{U^{[\Box]})}}{N}\,\mathcal{U}^{[+]}$ and
$\mathcal{U}^{[\Box]}\,\mathcal{U}^{[+]}$, respectively.

Here we will consider only the T-odd part of the contribution, the
so call gluonic-pole contribution, since we are interested in the
contribution of products of two $h^\p_1$. As shown in
Ref.~\cite{bacchtta05,bomhof07}, after weighting by the transverse
momentum, the $T$-odd part of each basic soft correlator will appear
as the T-odd distribution function (in our case the T-odd functions
is $h^{\p (1)}_1$) multiplied by a prefactor. For example, the
correlator $\Phi^{[(\Box)+]}$ will give a factor 1, while
$\Phi^{[\Box +]}$ will contribute a factor 3. Therefore, the factor
contributed by the product of two $h_1^\p$ is:
\begin{eqnarray}
&&\left(\frac{2N^2}{N^2-1}\times 1 \times
1-\frac{N^2+1}{N^2-1}\times 3\times 3\right) \nonumber \\
&=& -\frac{7N^2+9}{N^2-1}.\label{gpf}
\end{eqnarray}
We emphasize that this factor is the result coming from the $T$-odd
distributions due to their complicated gauge-link structure,
which cannot be predicted by the generalized parton model. After
considering the color factor from the hard process, we get
$C_{qq}$ as:
\begin{eqnarray}
C_{qq}&=&-\frac{N^2-1}{4N^3} \times \left (-\frac{7N^2+9}{N^2-1} \right )\nonumber \\
&=& \frac{7N^2+9}{4N^3}.
\end{eqnarray}

Therefore the final result of the $\cos \delta \phi$-moment shown in
Eq.~(\ref{csqq3}) is
\begin{eqnarray}
\frac{d\sigma^{W}[J_q \, J_q]}{d P_{\p}^2  d\eta_1 d\eta_2} & = &
\frac{4 \pi \alpha_s^2 }{s \boldsymbol{P}_\p^2} \frac{7N^2+9}{N^3}
\,y(1-y) \nonumber \\
& \times &\sum_q  h_1^{\p (1) q}(x_1) h_1^{\p (1) q}(x_2).
\label{csqq2}
\end{eqnarray}
We then can define a subprocess $\cos \delta \phi$ asymmetry in
$H_A+H_B \to J_q + J_q +X$ processes as
\begin{widetext}
\begin{eqnarray}
R_{qq}= \frac{d\sigma^{\mathcal{W}}[J_q \, J_q]/d P_{\p}^2 d\eta_1
d\eta_2}{d\sigma^{O}[J_q \, J_q]/d P_{\p}^2 d\eta_1 d\eta_2 } =
\frac{4 (7N^2+9) \sum_q  h_1^{\p (1) q}(x_1) h_1^{\p (1) q}(x_2)}{
\left \{\frac{N^3-N}{2}\left
(\frac{1+(1-y)^2}{y^2}+\frac{1+y^2}{(1-y)^2}\right
)-(N^2-1)\frac{1}{y(1-y)} \right \} \sum_q
 f_1^q \,(x_1)\,f_1^q \,(x_2)}.\label{rqq1}
\end{eqnarray}
\end{widetext}
The main result given by this asymmetry is that there is an
additional factor $-\frac{7N^2+9}{N^2-1}$, as shown in (\ref{gpf}),
apart from the one obtained in the standard generalized parton
model. The size of this gluonic pole factor is -9 for $N=3$, which
is quite large and will greatly enhance the asymmetry. Moreover,
comparing to the result from generalized parton model, the sign of
the asymmetry is reversed by this gluonic pole factor.

\subsection{\bf Process $H_A + H_B \to J_q+J_{\bar{\,q}}+X$}

Now we will consider the process $H_A + H_B \to
J_q+J_{\,\bar{q}}+X$, where the partonic subprocesses contributing
include $q \, \bar{q} \to q \, \bar{q}$ and $g \, g \to q \, \bar{q}
$, as shown in Figs. \ref{qqbarfig} and \ref{qqbarfig2},
respectively. The diagram in \ref{qqbarfig}a is also present for $q
\, \bar{q} \to q^\prime \, \bar{q}^\prime$, which will contribute to
$J_q+J_{\bar{q}}$ di-jet production. As before, the angular
independent cross-section of $H_A + H_B \to J_q+J_{\bar{q}}+X$
process can be readily written as
\begin{widetext}
\ba
 \frac{d\sigma^O[J_q \, J_{\bar{q}}]}{d^2 \boldsymbol{P}_{1\p}d^2 \boldsymbol{P}_{2\p} d\eta_1
d\eta_2}&=& \frac{\alpha_s^2}{s\hat{s}}\int d^2 \boldsymbol{k}_{1\p}
d^2 \boldsymbol{k}_{2\p} \delta^2
(\boldsymbol{k}_{1\p}+\boldsymbol{k}_{2\p}-\boldsymbol{P}_{1\p}-\boldsymbol{P}_{2\p})
\left \{ \left (\frac{N^2-1}{N^2}\left
(\frac{1+(1-y)^2}{2y^2}+(1-y)^2+y^2\right ) \right.
\right.\nonumber \\
& & \left.  +\frac{(N^2-1)}{N^3}\frac{(1-y)^2}{y} \right ) \sum_q
(f_1^q \,(x_1,\boldsymbol{k}_{1\p}^2)\,
f_1^{\bar{q}}\,(x_2,\boldsymbol{k}_{2\p}^2)+(x_1 \leftrightarrow
x_2))
\nonumber \\
& & \left. + \left
(\frac{1}{2N}\frac{y^2+(1-y)^2}{y(1-y)}-\frac{N}{N^2-1}y^2(1-y)^2
\right )f_1^g \,(x_1,\boldsymbol{k}_{1\p}^2) f_1^g
\,(x_2,\boldsymbol{k}_{2\p}^2)\right \} ,
 \ea
 \end{widetext}
 and the corresponding $\boldsymbol{P}_{2\p}$-integrated cross-section is given by
 \begin{widetext}
\ba \frac{d\sigma^O[J_q \, J_{\bar{q}}]}{d P_{\p}^2 d\eta_1 d\eta_2}
& = & \frac{\pi \alpha_s^2}{s\hat{s}} \left \{ \left
(\frac{N^2-1}{N^2}\left (\frac{1+(1-y)^2}{2y^2}+1-2y+2y^2\right
)+\frac{(N^2-1)}{N^3}\frac{(1-y)^2}{y}\right) \sum_q (f_1^q
\,(x_1)\, f_1^{\bar{q}}\,(x_2)
\right. \nonumber \\
 &  & \left.
+(x_1 \leftrightarrow x_2)) +\left
(\frac{1}{2N}\frac{y^2+(1-y)^2}{y(1-y)}-\frac{N}{N^2-1}y(1-y) \right
) f_1^g \,(x_1) f_1^g \,(x_2) \right \},\label{csqqbar}
 \ea
\end{widetext}

\begin{figure}

\begin{center}
\scalebox{0.50}{
\includegraphics[10pt,15pt][288pt,207pt]{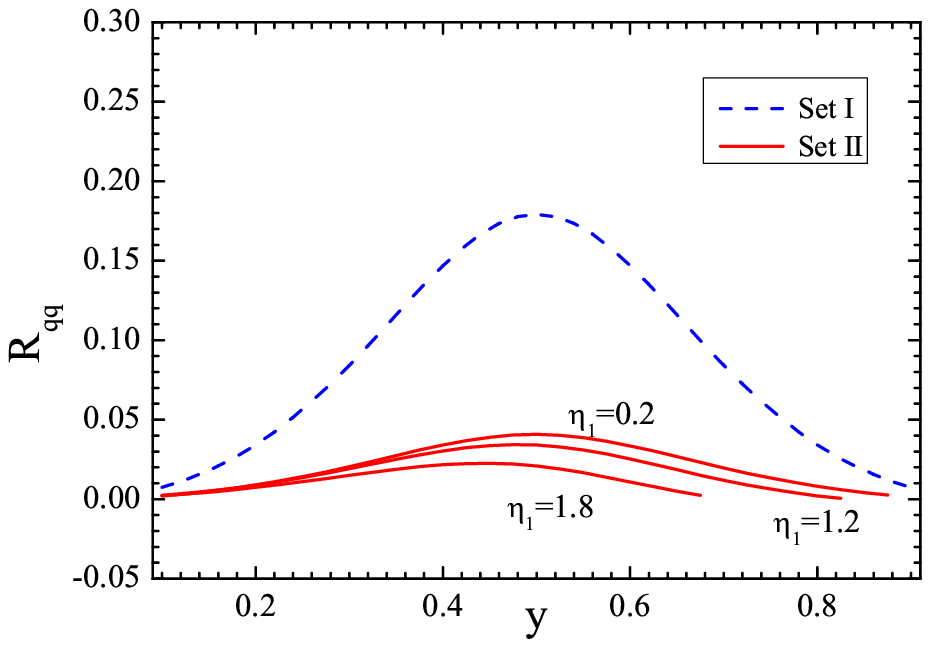}
\includegraphics[34pt,15pt][282pt,208pt]{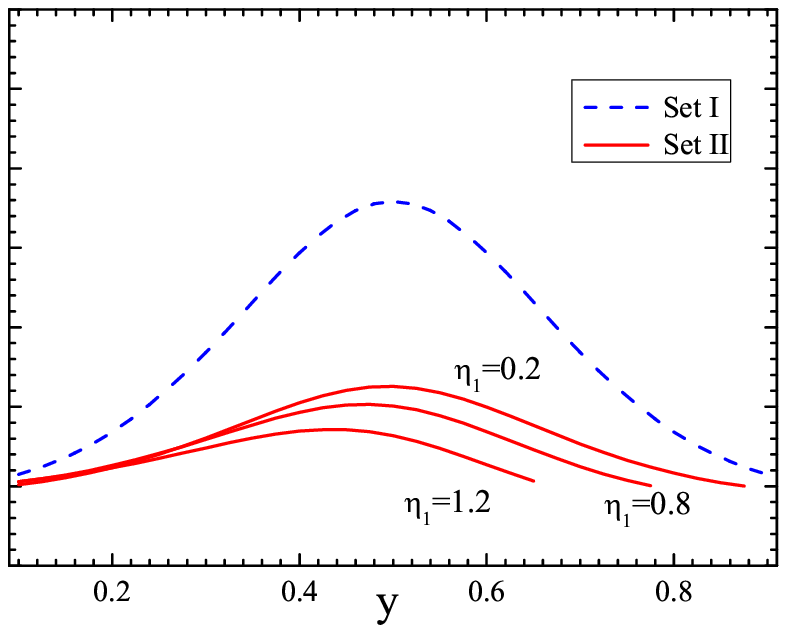}}\\
\scalebox{0.50}{
\includegraphics[-13pt,15pt][265pt,207pt]{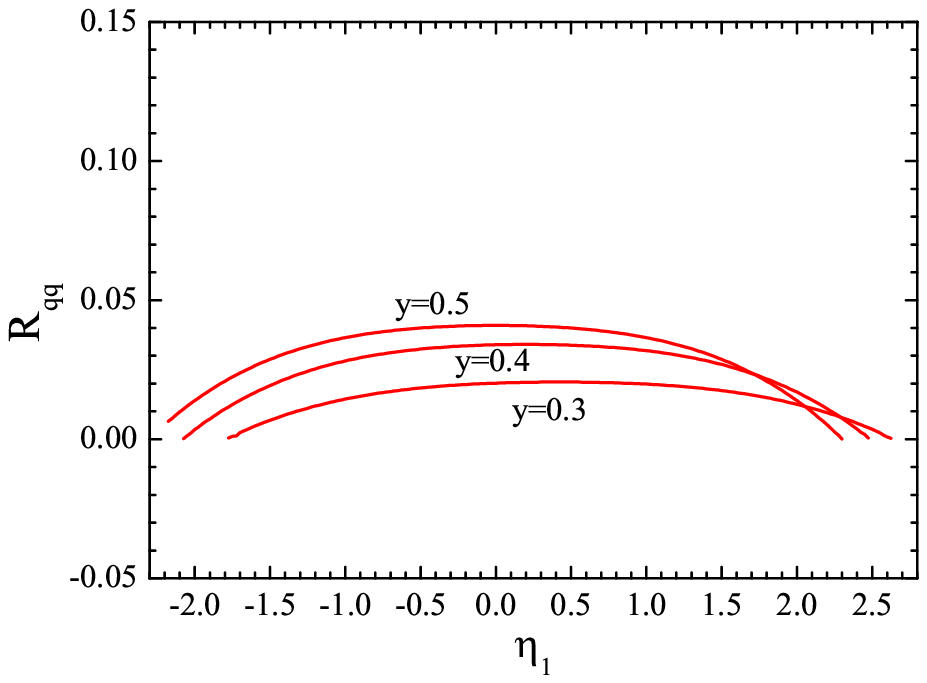}
\includegraphics[-3pt,12pt][286pt,208pt]{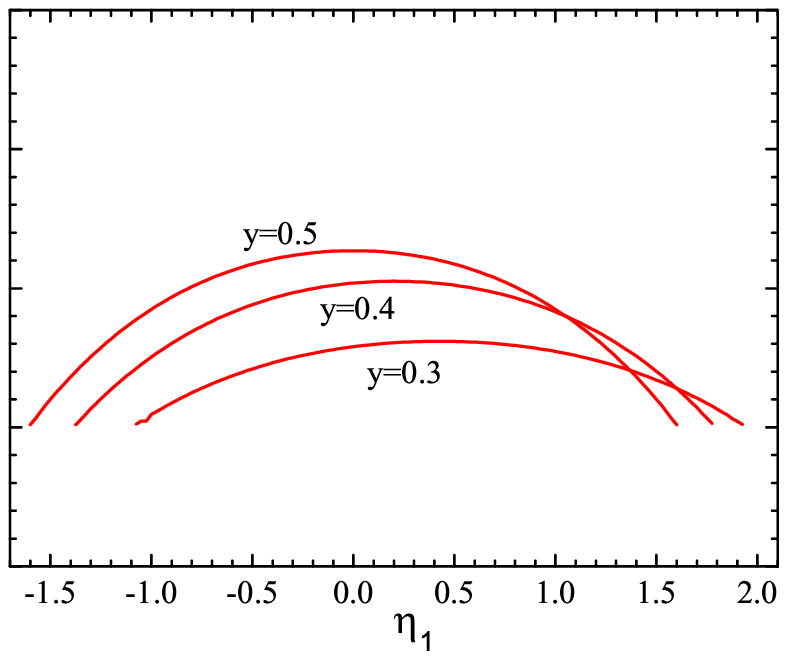}}\caption{\small
 The subprocess $\cos \delta \phi$ asymmetry $R_{qq}$ in di-jet
production $H_A + H_B \to J_q+J_q+X$ at $\sqrt{s}=200 \textrm{GeV}$,
for $P_\p= 10$ GeV (left column) and 20 GeV (right column). The
dashed line and solid lines are the results from taking Boer-Mulders
functions of Set I and II, respectively. The upper and lower panels
show the asymmetry vs $y$ and $\eta_1$, respectively.}
\label{rqqfig}
\end{center}

\end{figure}

\begin{figure}

\begin{center}
\scalebox{0.75}{
\includegraphics[10pt,15pt][288pt,207pt]{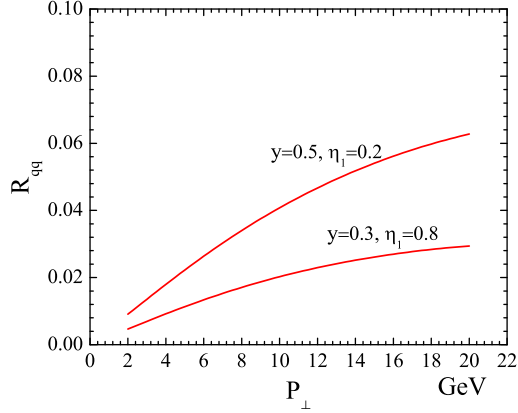}}
\caption{\small The $P_\p$-dependent subprocess $\cos \delta \phi$
asymmetry $R_{qq}$ in di-jet production $H_A + H_B \to J_q+J_q+X$ at
$\sqrt{s}=200 \textrm{GeV}$ calculated from the Boer-Mulders
functions in Set II.} \label{rqqfigpt}
\end{center}

\end{figure}

Again, since $h_1^\p$ is a chiral-odd function which flips the
helicity of the initial and final partons, the hard diagrams that
can contribute to the angular asymmetry in $H_A + H_B \to J_q +
J_{\bar{q}}$ are given in Fig.~\ref{qqbarfig}a, Fig.~\ref{qqbarfig}c
and Fig.~\ref{qqbarfig}d. Following a similar procedure as that used
in the previous subsection, we can give the $\cos \delta
\phi$-moment of $H_A + H_B \to J_q + J_{\bar{q}}$ using the
weighting function $\mathcal{W}=\frac{ P_\p^2 \cos \delta
\phi}{M^2}$, as follows
\begin{eqnarray}
\frac{d\sigma^{\mathcal{W}}[J_q \, J_{\bar{q}}]}{d P_{\p}^2 d\eta_1
d\eta_2} &=& \int d \phi_1 d^2 \boldsymbol{P}_{2\p}  \mathcal{W}
\frac{d\sigma^A[J_q
\, J_{\bar{q}}]}{d^2 \boldsymbol{P}_{1\p}d^2 \boldsymbol{P}_{2\p} d\eta_1 d\eta_2} \nonumber \\
&=& \frac{16 \pi  \alpha_s^2 }{s P_\p^2} \, \left (C_{q\bar{q}} \,
y^2(1-y)^2+ C_{q\bar{q}}^\prime \, y^2(1-y)\right)\nonumber \\
& \times &\sum_q \, (\,h_1^{\p,(1),q}(x_1) \,
h_1^{\p,(1),\bar{q}}(x_2) + (x_1 \leftrightarrow x_2 ) ),\nonumber
\\
\label{csqqbar2}
\end{eqnarray}
The factors $C_{q\bar{q}}$ and $C_{q\bar{q}}^\prime$ can be
calculated by combining the contribution of the hard subprocess and
the gluonic pole factor. The term with $C_{q\bar{q}}$ is the result
coming from the hard process in Fig.~\ref{qqbarfig}a, and the color
factor of this process is $\frac{N^2-1}{4N^2}$. The term with
$C_{q\bar{q}}^\prime$ is the result from the hard processes
Fig.~\ref{qqbarfig}c and Fig.~\ref{qqbarfig}d. The color factor of
these processes is $-\frac{N^2-1}{4N^3}$. The gluonic pole factors
for those processes can be calculated from the corresponding
gauge-links. For the diagrams in Fig.~\ref{qqbarfig}a, c and d, the
gauge-links are equivalent, and given by:
\begin{eqnarray}
\mathcal{U}_{q\bar{q}}^{[ss^*]} &=& \mathcal{U}_{q\bar{q}}^{[st^*]}
= \mathcal{U}_{q\bar{q}}^{[ts^*]} =
\frac{N^2}{N^2-1}\Phi^{[(\Box)+]}\nonumber\\
& &-\frac{1}{N^2-1}\,\Phi^{[-]}. \label{wl2}
\end{eqnarray}
The convolution of the soft parts in the decomposition of two color
flows is therefore
\begin{eqnarray}
\Phi^{[\mathcal{U}]}\otimes \Phi^{[\mathcal{U}^\dag]} & \sim &
\frac{2N^2}{N^2-1}\Phi^{[(\Box)+]}\otimes \Phi^{[(\Box)+^\dag]}
\nonumber \\
&&-\frac{1}{N^2-1}\Phi^{[-]}\otimes \Phi^{[-^\dag]},
\end{eqnarray}
Therefore the factor contributed by the product of $h_1^\p$ and
$\bar{h}_1^\p$ is calculated as:
\begin{eqnarray}
\frac{N^2}{N^2-1}\times 1 \times 1-\frac{1}{N^2-1}\times (-1)\times
(-1) = 1.\label{qqbargpf}
\end{eqnarray}

Thus we conclude that $C_{q\bar{q}}=\frac{N^2-1}{4N^2}$ and
$C_{q\bar{q}}^\prime=-\frac{N^2-1}{4N^3}$. Unlike in the case of
$J_q + J_q$ production, this result shows that in the case of $J_q +
J_{\bar{q}}$ production there is no color factor enhancement for the
azimuthal angle dependent cross-section, being therefore equivalent
to the generalized parton model result. Thus we can give the
subprocess $\cos \delta \phi$ asymmetry of $J_q + J_{\,\bar{q}}$ as
the ratio of Eq.~(\ref{csqqbar2}) and (\ref{csqqbar}):
\begin{eqnarray}
&& \hspace{-0.5cm} R_{q\bar{q}}= \frac{d\sigma^{\mathcal{W}}[J_q \,
J_{\bar{q}}]/d P_{\p}^2 d\eta_1 d\eta_2}{d\sigma^{O}[J_q \,
J_{\bar{q}}]/d P_{\p}^2 d\eta_1 d\eta_2 }= \nonumber \\
&& \hspace{-0.5cm} \frac{ A(y) \sum_q  \, (\,h_1^{\p,(1),q}(x_1) \,
h_1^{\p,(1),\bar{q}}(x_2) + (x_1 \leftrightarrow x_2 ) )}{ B(y)
\sum_q (f_1^q \,(x_1)\, f_1^{\bar{q}}\,(x_2) +(x_1 \leftrightarrow
x_2)) +C(y) f_1^g \,(x_1) f_1^g \,(x_2) },\label{rqqbar}
\end{eqnarray}
where
\begin{eqnarray}
A(y)&=&4 \left ( \, y(1-y)- \frac{y}{N}\right),\\
B(y)&=&  \frac{1+(1-y)^2}{2y^2}+(1-y)^2+y^2
\nonumber \\
& & +\frac{1}{N}\frac{(1-y)^2}{y}, \\
C(y)&=&\left
(\frac{N^2-1}{2N^3}\frac{y^2+(1-y)^2}{y(1-y)}-\frac{1}{N}y(1-y)
\right ).
\end{eqnarray}

\section{Numerical Results}

In the previous section we have calculated the theoretical result
for the azimuthal angle dependence of di-jet production in
unpolarized hadron scattering, and we showed that the azimuthal
angle dependent part of the cross-section can be separated from the
azimuthal angle independent part by taking a $\cos \delta
\phi$-moment. We defined the $\cos \delta \phi$ asymmetry as the
ratio of $\cos \delta \phi$-moment and the azimuthal angle
independent cross-section. It is interesting to study wether the
asymmetry could be accessed in hadron colliders. The Relativistic
Heavy Ion Collider (RHIC) at BNL is running the polarized di-jet
production process $p^\uparrow + p \to J_1 + J_2
+X$~\cite{rhicdijet}, with the main purpose of measuring the
transverse single spin asymmetry. The process can also be used to
access the unpolarized di-jet production by averaging the
spin-dependent data. We therefore calculate the  subprocess
asymmetry $R_{qq}$ given in Eq.~(\ref{rqq1}) at RHIC energy
($\sqrt{s}=200$ GeV), and show the result in Fig.~\ref{rqqfig}. In
the calculation we take $N=3$ and apply two sets of Boer-Mulders
functions. Set I is the Boer-Mulders functions which saturate the
positivity bound:
\begin{eqnarray}
h_1^{\p(1)q} (x) \le \frac{\langle |\boldsymbol{k}_\p| \rangle}{2M}
f_1^q(x),
\end{eqnarray}
with $\langle |\boldsymbol{k}_\p| \rangle = 0.44$
GeV~\cite{anselmino05a}. Therefore this Set Boer-Mulders function
will give the maximum bound of the asymmetries. In Set II we apply
the Boer-Mulders functions which are extracted in Ref.~\cite{bing08}
from unpolarized $p+D$ Drell-Yan data~\cite{e866} measured by the
FNAL-E866/NuSea Collaboration. As shown in that paper, the
Boer-Mulders functions in Set II can successfully reproduce the
$\cos 2 \phi$ asymmetry in the unpolarized $p+D$ Drell-Yan process.
So far the information on these functions is limited, therefore we
adopt these two sets of Boer-Mulders functions to perform estimates
on the $\cos \delta \phi$ asymmetries. For the unpolarized parton
distributions appearing in the denominators of Eqs.~(\ref{rqq1}) and
(\ref{rqqbar}) adopt the MRST2001 (LO set) parametrization
~\cite{mrst}. The results for $R_{qq}$ from sets I and II are
plotted with dashed and solid lines, respectively. The upper panel
of Fig.~\ref{rqqfig} shows the subprocess asymmetries vs $y$ at
$P_\p = 10 \, \textrm{GeV}$ (left) and $P_\p=20\,\textrm{GeV}$
(right) respectively, while the lower panel shows the subprocess
asymmetries vs $\eta_1$ at same $P_\p$. In Fig.~\ref{rqqfigpt} we
also show $R_{qq}$ vs $P_\p$ which is calculated from the
Boer-Mulders functions in Set II. The result from Set I, which can
be understood as the upper limit of the asymmetry, can be more than
$10\%$ in size. The result from Set II is about several percent.
which is still quite sizable. The main reason of this result is the
enhancement factor of -9 on the asymmetry from T-odd distributions
in hadron scattering.

\begin{figure}

\begin{center}
\scalebox{0.35}{
\includegraphics{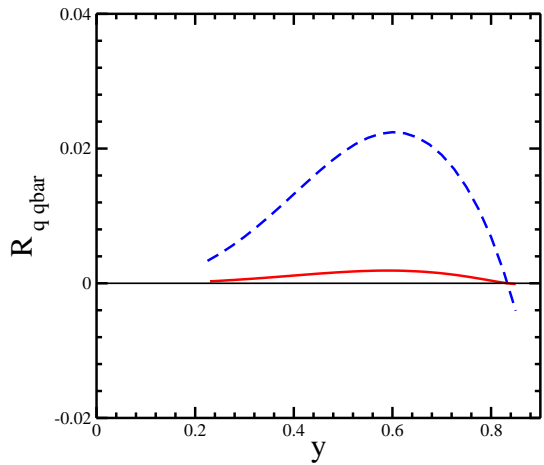}}
\caption{\small The subprocess $\cos \delta \phi$ asymmetry $R_{q
\bar{q}}$ in di-jet production $H_A + H_B \to J_q+J_{\,\bar{q}}+X$
at $\sqrt{s} =200$ GeV for $P_\p= 20$ GeV. The dashed line and solid
lines are the results from Boer-Mulders functions of Set I and II,
respectively.} \label{rqqbarfig}
\end{center}

\end{figure}

We also calculate the subprocess $\cos \delta \phi$ asymmetry of
$J_q + J_{\,\bar{q}}$ production given in Eq.~(\ref{rqqbar}) at
$\sqrt{s}=200$ GeV, and show $y$-dependent curves at $P_\p=20
\,\textrm{GeV}$ and $\eta_1=0.2$ in Fig.~\ref{rqqbarfig}. Our
calculation indicates a subprocess smaller asymmetry for $J_q +
J_{\,\bar{q}}$ production than that for $J_q + J_q$ production. This
can be understood from the fact that there is no color factor
enhancement for the azimuthal angle dependent cross-section of $J_q
+ J_{\,\bar{q}}$ production, and there is a large gluon-gluon
scattering contribution in the denominator of Eq.~(\ref{rqqbar}).

\begin{figure}
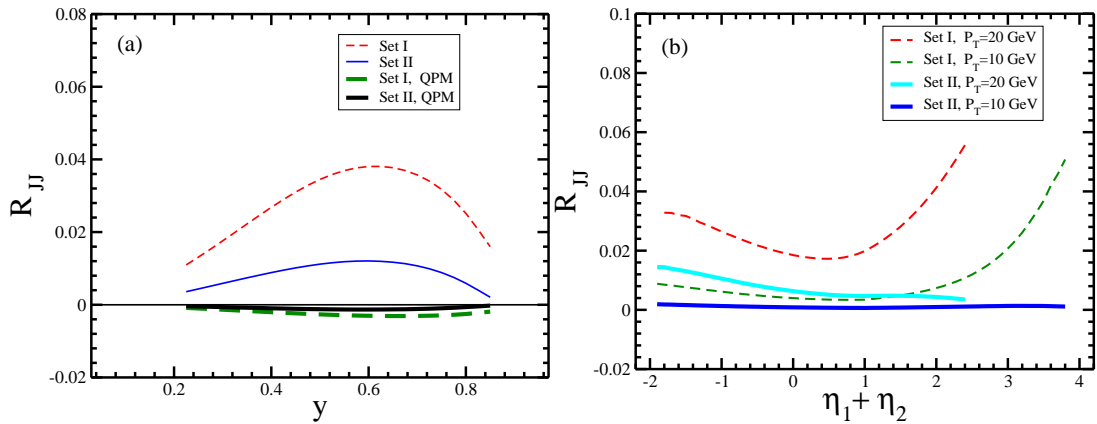


\begin{center}
\scalebox{0.35}{
\includegraphics{dijettotal.eps}
\includegraphics{sumeta.eps}} \caption{\small (a), The $y$-dependent $\cos \delta \phi$
asymmetry $R_{JJ}$ in di-jet production $H_A + H_B \to J_1+J_2+X$ at
RHIC for $P_\p= 20$ GeV. The dashed line and solid lines are the
results from Boer-Mulders functions of Set I and II, respectively.
We also show by thick lines the result of $R_{J J}$ from the
conventional generalized parton model (GPM) calculation where
gluonic pole contributions are not considered. (b), The $\cos \delta
\phi$ asymmetry $R_{JJ}$ vs $\eta_1+\eta_2$ in di-jet production
$H_A + H_B \to J_1+J_2+X$ at RHIC for $P_\p= 10$ GeV (thick lines)
and $P_\p= 20$ GeV (thin lines) from Boer-Mulders functions of Set I
(thin lines) and II (thick lines).} \label{rtotal}
\end{center}

\end{figure}

In a practical measurement of di-jet production, the flavor of the
jets usually is not measured. Hence the subprocess asymmetries shown
in Figs.~\ref{rqqfig}, \ref{rqqfigpt} and \ref{rqqbarfig} are not
measurable. The purpose to present those figures is to show how much
the contribution coming from each subprocess is. Therefore we
calculate the $\cos \delta \phi$ asymmetry of any di-jet production
$R_{J J}$, which is measurable and defined as
\begin{eqnarray}
R_{J J} = \frac{\sum_{a,b= q,\bar{q},g} d\sigma^{\mathcal{W}}[J_a \,
J_b]/d P_{\p}^2 d\eta_1 d\eta_2}{\sum_{a,b= q,\bar{q},g}
d\sigma^{O}[J_a \, J_b]/d P_{\p}^2 d\eta_1 d\eta_2}. \label{rj1j2}
\end{eqnarray}
In this paper we assume that the partonic processes that contribute
to the azimuthal angular dependence of di-jet production are $q q
\to q q$ and $q \bar{q} \to q \bar{q}$. Therefore the numerator in
Eq.~(\ref{rj1j2}) is the sum of Eqs.~(\ref{csqq2}) and
(\ref{csqqbar2}), while the denominator contains the contributions
from Eqs.~(\ref{csqqo}), (\ref{csqqbar}) and from all other possible
processes, such as $q q^\prime \to q q^\prime$, $q g \to qg$, and so
on. In Fig.~\ref{rtotal} we show  by thin lines the asymmetry $R_{J
J}$ at RHIC for $P_\p= 20$ GeV vs $y$, from the two sets of
Boer-Mulders functions (solid and dashed lines for Set I and Set II
respectively). As a comparison, we also show by thick lines the
result of $R_{J J}$ from the conventional generalized parton model
(GPM) calculation where gluonic pole contributions are not
considered. In Fig.~\ref{rtotal}b, we show the asymmetry $R_{J J}$
at RHIC for $P_\p= 10$ GeV (thick lines) and $P_\p = 20$ GeV (thin
lines) vs $\eta_1 + \eta_2$. In the calculations we restrict $-1<
\eta_{1/2}< 2$ to consider the pseudorapidity acceptance of jet at
RHIC. The curves given in Fig.~\ref{rtotal}a indicate a sign
reversal of the asymmetries $R_{J J}$ between the results from the
two different approaches. This is due to the additional factor $-9$
in the azimuthal angular dependent part of $H_A + H_B \to
J_q+J_q+X$, coming from the multiple initial- and final-state
interactions. We remind that a similar sign reversal also happens in
the SSA in hadronic photon-jet production~\cite{bacchetta07}, which
is due to a similar effect. Therefore the measurement of the $\cos
\delta \phi$ asymmetry $R_{JJ}$ in unpolarized di-jet production at
RHIC, especially the enhancement of the size and the sign reverse of
the asymmetry, will provide further tests on the role of initial-
and final-state interactions as QCD dynamics of $T$-odd
distributions in hadron scattering. We would like point out that in
calculating $R_{JJ}$, we only considered the lowest order
contributions to parton scattering. Other contributions may affect
the size of the asymmetry, such as NLO corrections for the
denominator of Eq.~(\ref{rj1j2}). At large $P_\perp$ they are
further enhanced by threshold logarithms which are more pronounced
for gluonic channels. Therefore our results are estimates and
$R_{JJ}$ can be smaller. However we expect that those contributions
will not change the sign of the asymmetry.

\section{Summary}

In summary, we performed a study of the azimuthal angular dependence
of back-to-back di-jet production in unpolarized hadron scattering,
focusing our attention on the process $H_A+H_B \to J_q + J_q +X$,
where two quark jets have the same flavor, and on the process
$H_A+H_B \to J_q + J_{\bar{q}} +X$, where the di-jet comes from a
quark-antiquark pair. We find that in both processes there are an
azimuthal angle dependent cross-section generated by the product of
two $h_1^\p$ from each incident hadron. Using a weighting function
$\cos \delta \phi$, we can separate the azimuthal angle dependent
cross-section from the azimuthal angle independent one. We defined
the $\cos \delta \phi$ asymmetry of di-jet production as the ratio
between the $\cos \delta \phi$-moment and the azimuthal angle
independent cross-section. In the case of $J_q + J_q $ production,
due to the multiple initial- and final-state interactions within
hadron scattering, there is a color factor enhancement (which is -9
for $N=3$) in the asymmetry compared to the result from the standard
generalized parton model. No such enhancement appears in the case of
$J_q + J_{\bar{q}}$ production. We estimate the $\cos \delta \phi$
asymmetry of di-jet production at the RHIC energy, and find that the
subprocess asymmetry of $J_q + J_q$ production can reach up to 10\%
in maximum due to the large color factor enhancement. In contrast,
the subprocess asymmetry of $J_q + J_{\,\bar{q}}$ production is
smaller. We further find that the color factor enhancement in $J_q +
J_q$ production can increase the size and reverse the sign of the
total $\cos \delta \phi$ asymmetry $R_{JJ}$ in di-jet production.
Therefore it is feasible to perform the measurement on the di-jet
production in unpolarized hadron scattering at RHIC, which can
identify the color factor enhancement on the $\cos \delta \phi$
asymmetry in $J_q + J_q$ production. Furthermore, the study will
provide the opportunity for a better understanding on the role of
initial- and final-state interactions in hadron scattering and other
processes.

\bigskip

{\bf Acknowledgements} This work is supported by the PBCT project
ACT-028 ``Center of Subatomic Physics".

\end{document}